**Terahertz emission on the surface of a van der Waals magnet CrSiTe$_3$**

*Peng Suo[1], Wei Xia[2], Wenjie Zhang[1], Xiaoqing Zhu[1], Jibo Fu[1], Xian Lin[1], Zuanming Jin[1,3], Weimin Liu[2,3], Yanfeng Guo[2,3*] and Guohong Ma[1,3*]*

[1] Department of Physics, Shanghai University, Shanghai 200444, China

[2] School of Physical Science and Technology, ShanghaiTech University, Shanghai 201210, China

[3] STU & SIOM Joint Laboratory for superintense lasers and the applications, Shanghai 201210, China

[*] Corresponding authors: phymagh@shu.edu.cn (G. Ma) and

guoyf@shanghaitech.edu.cn (Y. Guo)

**Abstract:** The van der Waals magnet CrSiTe$_3$ (CST) has captured immense interest because it is capable of retaining the long-range ferromagnetic order even in its monolayer form, thus offering potential use in spintronic devices. Bulk CST crystal has inversion symmetry that is broken on the crystal surface. Here, by employing ultrafast terahertz (THz) emission spectroscopy and time resolved THz spectroscopy, the THz emission of the CST crystal was investigated, which shows a strong THz emission from the crystal surface under femtosecond (fs) pulse excitation at 800 nm. Theoretical analysis based on space symmetry of CST suggests the dominant role of shift current occurring on the surface with a thickness of a few quintuple layers in producing the THz emission, in consistence with the experimental observation that the emitted THz amplitude strongly depends on the azimuthal and pumping polarization angles. The present study offers a new efficient THz emitter as well as a better understanding of the nonlinear optical response of CST. It hopefully will open a window toward the investigation on the nonlinear optical response in the mono-/few-layer van der Waals crystals with low-dimensional magnetism.

**Keywords**: van der Waals crystal, low-dimensional magnetism, terahertz emission, shift photocurrent

The quasi two-dimensional (2D) van der Waals (vdW) magnets, such as Cr$X$Te$_3$ ($X$ = Si, Ge), display extraordinary physical properties arising from the complex magnetic exchanges. [1-4] One prominent virtue of this family of materials is that even in the atomically thin limit they still exhibit intrinsic long-range magnetic order which was previously thought to be strictly prohibited by thermal fluctuations in 2D systems with magnetic isotropy, thus offering opportunities to fundamentally study this peculiar low-dimensional magnetism and explore the potential use in spintronics. [5-7] Furthermore, the combination of the semiconductivity and magnetism in a single material provides ideal candidate for potential applications in optoelectronics. [8]

Most of the present attentions are paid on the magnetic properties of Cr$X$Te$_3$, [1-6,9-12] while the optical properties, especially the nonlinear optical response, remain nearly uninvestigated. [11,13-16] In fact, bulk Cr$X$Te$_3$ crystal shows $C_{3i}$ point group, in which the second order nonlinear effect vanishes due to the inversion symmetry of the crystal. It is noted that the Cr$X$Te$_3$ is layered structure with unit cell consisting of three quintuple layers (QLs), and distance between neighboring layers is 0.33 nm with the thickness of each QL of 0.35 nm, [11,14]. The inversion symmetry of the crystal is broken on the surface with thickness of a few QLs, and the second order nonlinear effect subsequently could appear on the surface of the crystal. In addition, as a semiconductor with narrow band gap, Cr$X$Te$_3$ may exhibit optical response intimately related to the intrinsic properties. To guide the light into this shadow area, we performed study on the nonlinear optical response of the CrSiTe$_3$ single crystal at room temperature by employing THz emission spectroscopy [17-20] combined with optical pump and THz probe (OPTP) spectroscopy. [21-22] Interestingly, a broad band THz radiation upon fs pulse illumination with photoenergy above bandgap is observed, which was confirmed to arise from the surface linearly photogalvanic effect (LPGE), i.e., shift current under linearly polarized photoexcitation. The strong THz radiation manifests the photoinduced large shift current, which arises from charge transfer process from $Te^{2-}$ to $Cr^{3+}$ ions. [23] In addition, we also observed weak THz radiation under circularly polarized excitation, which arises from the injection current produced via circularly photogalvanic effect (CPGE) on the surface of the CST crystal. Moreover, optical pump and THz probe spectroscopy measurements also reveal two relaxation processes of the photocarrier dynamics, i.e. the electron-phonon coupling and the surface defect trapping processes. Both of the two relaxation processes are too slow to contribute to the ultrafast THz emission, and the strong THz emission only takes place during the photoexcitation.

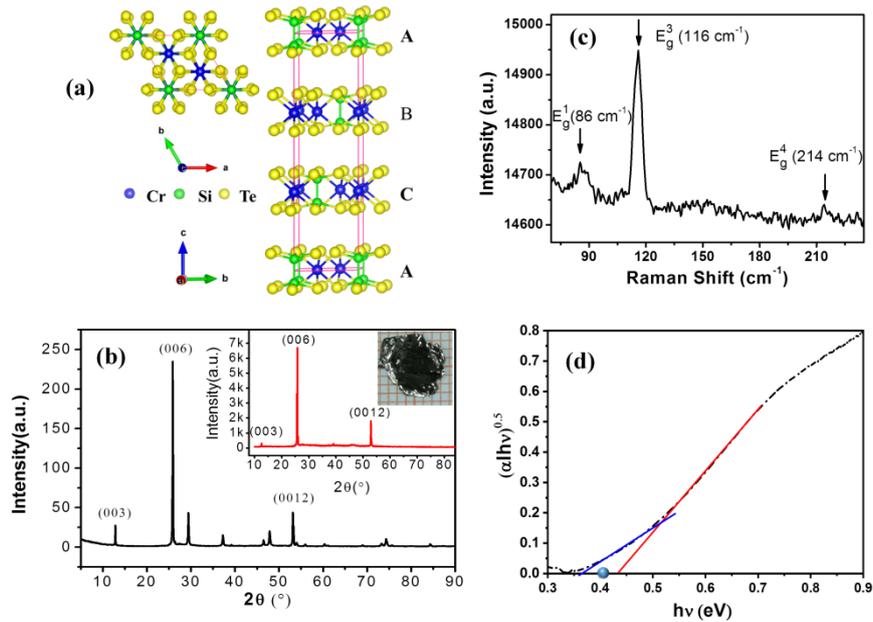

**Figure 1**. **Structure, quality and band gap of CST single crystal.** (a) Schematic crystal structures seen from different orientations (left: view from (*ab*) plane, right: view from (*bc*) plane). (b) Room temperature X-ray diffraction pattern of CST powder. The inset presents the XRD pattern and the picture of the *c*-cut CST single crystal. (c) Raman spectroscopy of CST single crystal with laser wavelength of 514.5 nm. (d) $(\alpha lh\nu)^{0.5}$ plotted against the photo energy $h\nu$. The interceptions of the red and blue lines with the horizontal axis give the sizes for the indirect band gaps ranged from 0.435 to 0.370 eV.

Bulk CST crystallizes into a layered structure with each unit cell being constructed by three CST layers stacking in an *ABC* sequence, in which the short Si-Si bonds form Si pairs and hence $Si_2Te_6$ ethane-like groups. The Cr ions locate at the centers of slightly distorted octahedron of Te atoms. The Cr ions and Si pairs, in a 2:1 ratio, form the octahedral site plane and are sandwiched between Te planes. The crystal structure of CST is illustrated in Figure 1(a) seen along different orientations; a quintuple layer can be clearly seen in each layered structure. Figure 1(b) presents the room temperature powder x-ray diffraction (XRD) pattern that could be indexed on the basis of $R\bar{3}$ space group. [11] The crystal was cut along the *ab*-plane, i.e. *c*-cut, for later measurements, shown by the inserted picture in Figure 1(b). The inset also shows the XRD pattern collected on the surface of *c*-cut crystal, indicating good crystallinity of our crystals. Figure 1(c) presents the Raman scattering of the CST single crystal at room temperature with cross polarization configuration, revealing the three Raman peaks locating at 86 cm$^{-1}$, 116 cm$^{-1}$ and 145 cm$^{-1}$ which are assigned as $E_g^1$, $E_g^3$ and $E_g^4$ modes, respectively. The asymmetry profile of the most pronounced feature of $E_g^3$ at low temperature evidences the spin-phonon coupling in the crystal. [14] The absorbance, $\alpha l$, was obtained by measuring the UV-visible absorption spectrum at room temperature. Figure 1(d) plots the $(\alpha h\nu)^{0.5}$ against $h\nu$, where $\alpha l$, $h$, and $\nu$ are the absorbance, Planck constant, and frequency, respectively. The band gap energy, $E_g$, is determined to be about 0.40±0.03 eV, agreeing well with the calculated value. [11] The band structure of CST presented in the Supplementary Information (SI) by Figure S1 was calculated by means of the first-principle method with VASP package, [24] which clearly shows an indirect gap semiconductor nature at room temperature with the gap of 0.43 eV and a direct band gap of 1.20 eV. From the analysis of the indirect band gap shown in Figure 1(d), the coupling phonon energy is determined as about 65 meV, which is much larger than that of fundamental phonon modes, suggesting the important role(s) of the overtone and/or combination modes in the optical transition. [11]

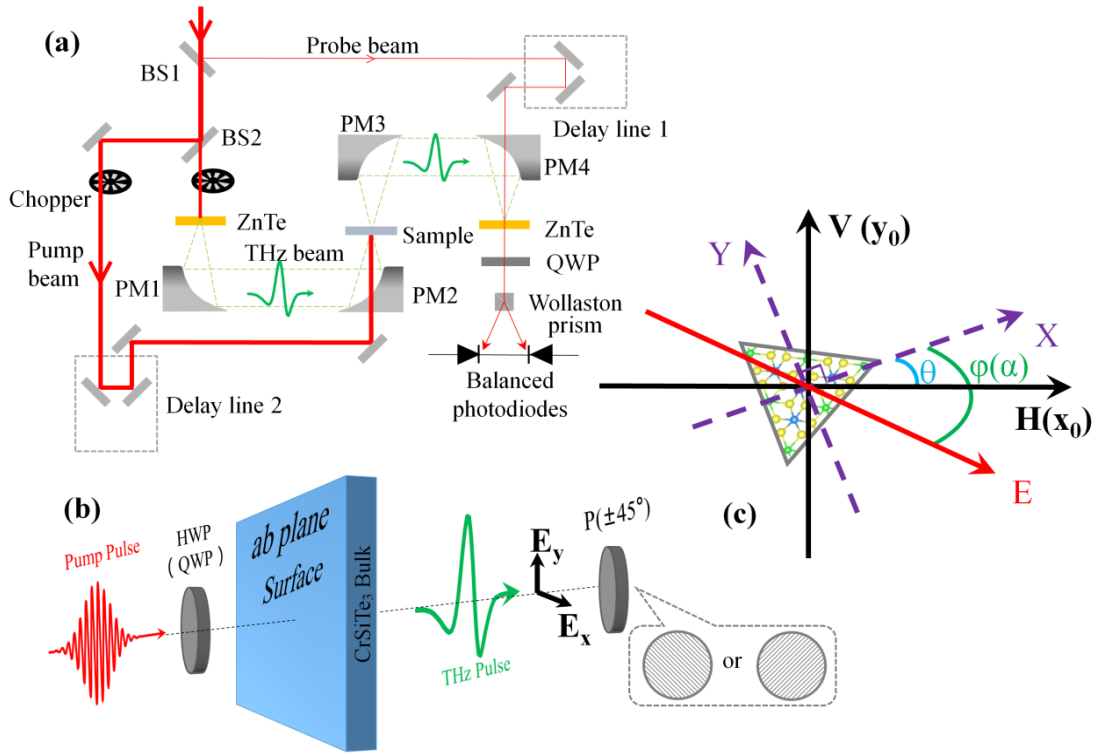

**Figure 2. Experimental setups.** (a) Schematics of experimental arrangements for optical pump and THz probe spectroscopy. (b) Illustration of experimental arrangement for THz emission

spectroscopy, HWP (QWP) and P denote the half-wave plate (quarter-wave plate) and THz polarizer, respectively. (c) Illustration of relative positions of laboratory coordinate system ($x_0$-$y_0$), sample coordinate system (x-y), as well as the direction of polarization (***E***) of the incident laser pulse, in which $\varphi(\alpha)$ denotes the azimuthal (polarization) angle.

Our experimental setups for OPTP and THz emission measurements are depicted schematically in Figures 2 (a) and (b), respectively. Figure 2(c) illustrates the relative position of laboratory coordinate system ($x_0$-$y_0$) ($x_0$ and $y_0$ are set along the horizontal (H) and vertical (V) directions, respectively), the sample coordinate system (x-y), and the polarization direction of incident laser pulse. The THz beam was enclosed and purged with dry nitrogen to avoid water vapor absorption. All measurements were performed at room temperature unless we specifically noted. As shown in Figure S2.1 in SI, CST has a refractive index of 3.17 with negligible absorption to the investigated THz frequency; therefore the THz emission spectroscopy with transmission configuration is applied, which makes the data analysis much simpler. We would like to note that the *c*-cut CST crystal shows isotropy in the THz transmission in the *ab*-plane, and is almost independent of the temperature in the investigated THz frequency, which are presented respectively in Figure S2.2 and Figure S2.3 in SI.

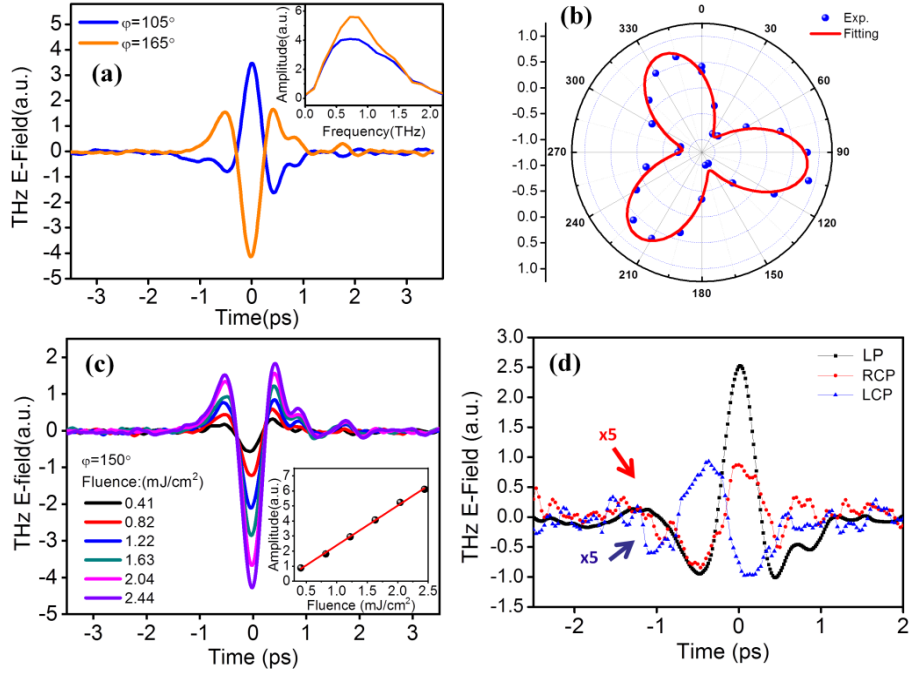

**Figure 3. THz radiation of *c*-cut CST single crystal induced by fs laser pulse.** (a) THz emission spectra in time domain for two typical azimuthal angles, and the corresponding Fourier transformation spectra (inset of (a)). (b) The amplitude of emitted THz electric field dependence on the azimuthal angle φ, the solid line is the fitting curve with sine function. (c) The emitted THz waveforms in time domain with various pump fluence, and the inset shows the peak-to-peak magnitude of THz radiation with respect to the pump fluence. The solid line is linear fitting. (d) THz emission spectra in time domain under 800 nm pump with linear (black), right-handed circular (red) and left-handed circular (blue) polarization, the red and blue curves are magnified by 5-time for comparison.

Figure 3 displays the main results of THz radiation of the *c*-cut CST single crystal induced by fs laser pulse at central wavelength of 800 nm. The absorption coefficient of CST to 800 nm laser is about $1.6\times10^5$ cm$^{-1}$, [11] indicating an optical penetration depth of about 62.5 nm and the surface effect plays a dominated role on the optical pump and THz probe spectroscopy, which will be discussed later. Figure 3(a) shows the fs pulse induced THz radiation for two selected azimuthal angles, i.e. $\varphi = 105°$ and 165°, respectively. The inset in Figure 3(a) is the Fourier transformation of the data shown in the main panel. The THz emission has a broad bandwidth covering from 0.1 to 2.0 THz. The polarity of THz emission reverses its sign when the azimuthal angle changes 60°. Figure 3(b) plots the azimuthal angle dependent amplitude of the generated THz electric field, it is clearly seen that the THz emission strength shows a triple symmetry, which shows a good consistence with the structural symmetry in *ab*-plane of the crystal. Figure 3(c) shows the pump fluence dependence of emitted THz waveforms at the azimuthal angle of 150°, revealing a good linear dependence on the pump fluence as is plotted in the inset of Fig. 3(c). Figure 3(d) compares the THz time domain spectra with linear and circular polarizations under identical fluence. It is seen that the magnitude of emitted THz radiation under circular polarization pumping is one-order of magnitude smaller than that of pumping with linear polarization. It is also noted that the THz radiation generated by fs pulse with right-handed circular polarization shows out-of-phase with that of left-handed circular polarization.

In order to evaluate the polarization of the THz radiation, a wire-grid polarizer P was placed between two parabolic mirrors, PM2 and PM3, as illustrated in Figure 2(b), in which P is aligned at polarization angle of ±45° with respect to the horizontal direction. The sum and difference of these two spectra at ±45° give the horizontal and vertical electric-field components, respectively. [25,26] Our experiment confirmed that the THz waveform obtained from the sum signals between +45° and -45° agrees well with the waveform measured directly at the horizontal component. Figure 4(a) presents the measured magnitudes of horizontal ($E_H$) and vertical ($E_V$) components of THz radiation with respect to the azimuthal angle $\varphi$. The amplitude of either $E_H$ or $E_V$ changes with $\varphi$ with a periodicity of 120°, consistent with the theoretical prediction based on nonlinear current model given later and harmonious with the triple symmetry in the *ab*-plane of CST. The solid lines in Figure 4(a) are fitting curves with sine function, which give the phase retardation between $E_H$ and $E_V$ of about 33.6° that is close to the theoretical value of $\pi/6$ given later. Figures 4 (b) and (c) present the three-dimensional (3D) plots of the generated THz polarization at $\varphi = 0°$ and 30°, respectively. The polarization of THz radiation changes from horizontal with $\varphi = 0°$ to vertical with $\varphi = 30°$, which does support the conclusion that $\varphi$ dependent amplitude of THz radiation shows three periodicities in the range of $\varphi = 0°$ to $2\pi$. More experimental data are presented in Figure S3 of the SI.

Next, we investigated the pump polarization dependence of THz radiation with the fixed angle, $\theta$. The pump polarization can be varied continuously by rotating a half-wave plate (HWP) inserted in the pump path. Figure 4 (d) shows the peak-to-peak amplitude of $E_H$ and $E_V$ components of THz radiation with respect to $\alpha$. The peaking amplitude oscillates with $\alpha$ in a periodicity of 180°. The solid lines are the fitting curves with sine function, which produces the phase retardation between $E_H$ and $E_V$ of about 51.1°, closed to the theoretical value of 45° as presented later. Figures 4(e) and (f) present the 3D plots of the polarization of THz radiation at $\alpha = 0°$ and 45°, respectively, showing that the polarization of THz radiation changes from horizontal with $\alpha = 0°$ to vertical with $\alpha = 45°$. The result is in support of that $\alpha$ dependent amplitude of THz radiation shows two periodicities in the range of $\alpha = 0$ to 2π. More experimental data about the 3D plots of the polarization of THz radiation with polarization angle changing from 0 ° to 300 ° are presented in Figure S4 of the SI.

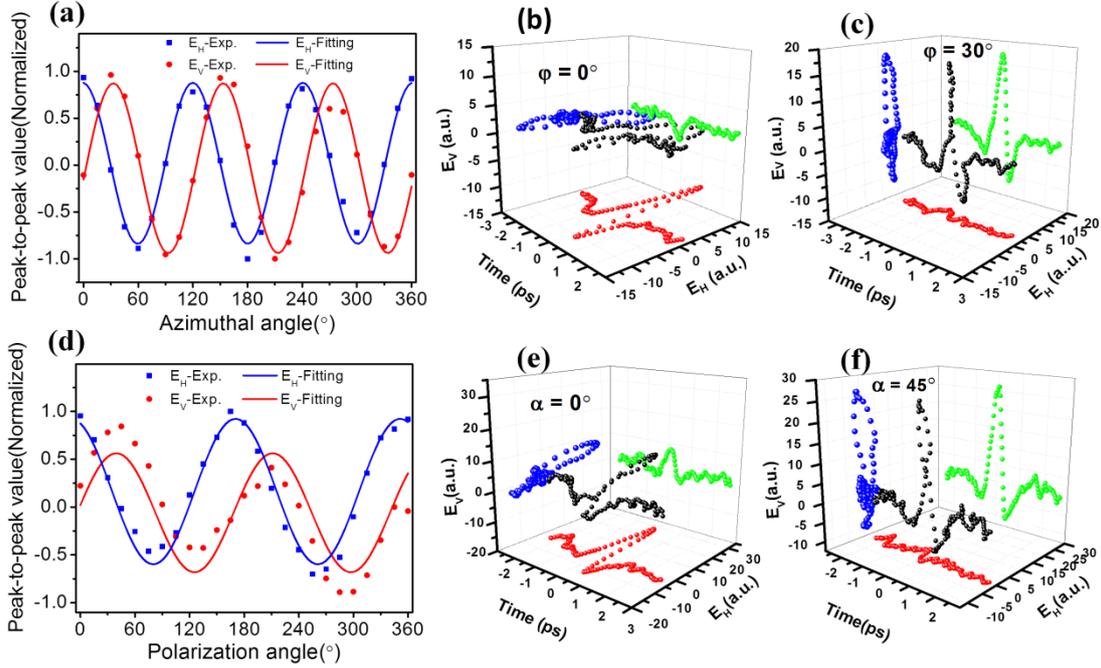

Figure 4. (a) Peak-to-peak amplitude of horizontal ($E_H$) and vertical ($E_V$) components of THz radiation with respect to φ from 0 to 2π. The solid lines are fitting curves with sine function. The 3D plots of THz radiation with two selected azimuthal angles are presented in (b) $\varphi$=0 ° and (c) 30 °. (d) Peak-to-peak amplitude of horizontal ($E_H$) and vertical ($E_V$) components of THz radiation with respect to $\alpha$ with fixed $\theta$. The solid lines are fitting curves with sine function. The 3D plots of THz radiation with two selected polarization angles are shown in (e) $\alpha$=0 ° and (f) 45 °.

In a non-magnetic semiconductor, THz radiation upon above band gap photoexcitation can be generally written as:

$$E_{THz} \propto \frac{\partial^2 P^{(OR)}}{\partial^2 t} + \frac{\partial J}{\partial t} \quad (1)$$

Here, $P^{(OR)}$ denotes the 2$^{nd}$-order polarization of optical rectification. The CST has inversion symmetry with space group $R\bar{3}$ ($C_{3i}$ point group), in which the 2$^{nd}$-order susceptibility is zero for bulk crystal. The THz transient may arise from the optical rectification on the surface, in which the inversion symmetry is broken. [27] To evaluate the relative importance of surface optical rectification in our case, we compared the emitted THz amplitude of our CST crystal with that of pump-transparent (110)-oriented ZnTe crystal under identical pump fluence. We assume that the THz pulse is emitted from the surface of CST crystal with a thickness of 2~3 nm (which is less than 3 QLs). After normalized with the thickness of ZnTe crystal, a conventional THz emitter, we find that the THz signal from CST surface is more than 600~900 times larger than that from ZnTe. Therefore, optical rectification effect on the surface of CST is expected to contribute negligibly, and the current term in the Eq. (1) is considered to make dominated contribution to the THz radiation.

For the above band gap photoexcitation, the ultrafast photocurrent are generally consisted of drift current ($J_{dri}$), diffusion current ($J_{dif}$), shift current ($J_{sh}$) and injection current ($J_{inj}$), i.e. $J=J_{dri}+J_{dif}+J_{sh}+J_{inj}$. [28-32] The drift current produced by surface depletion field and diffusion current induced by the photo-Dember effect are out of plane, perpendicular to the crystal surface, and the generated THz radiation consequently should be irrelevant to the azimuthal angle of the sample. In addition, considering that the pump pulse is normal incident onto the freshly cleavage surface in our experiment, the emitted THz radiation due to the drift current and diffusion current is expected to propagate along the surface, which can not be detected according to our experimental configuration. Moreover, as shown in the inset to Figure 3(c) the THz radiation amplitude varies linearly with the optical pump intensity, indicating the drift current and diffusion current are negligible to influence the current amplitude. At last, photo drag effect (PDE) can also generate nonlinear photocurrent, leading to the THz radiation.[29,33-34] In fact, the PDE is a fourth rank nonlinear effect, which arises from the momentum transfer from the incident photons to the electrons near the surface in the penetration depth.[33-34] The photogenerated charge carriers in the CST crystal will move along the direction of the incident light and generate transient photocurrent. Considering the normal incidence of light in our case, the photocurrent due to the photo drag effect is normal to the sample surface, and the induced THz radiation is expected to propagate along the sample surface, which cannot be detected by our experimental configuration. Another significant feature of the PDE is that the photocurrent changes its sign by reversing the wave vector. We have obtained almost in-phase THz radiation with identical amplitude within our experimental error by photoexcitation of the front and back sides of the sample. Therefore, the strong THz emission in the present study can only arise from the injection current and shift current.

The bulk CST is denoted as $C_{3i}$ point group, which shows inversion symmetry, and the inversion symmetry is broken at the surface/interface, in which the $C_3$ point group is applicable. As illustrated in Figure 2(c), the x-y system is assigned as a sample coordinate system, in which the angle $\theta$ is the included angle between x and $x_0$. The $\boldsymbol{E}$ is the polarization direction of incident optical pulse and the $\varphi$ ($\alpha$) denotes the angle between x and $\boldsymbol{E}$. The fs pulse propagating along z($c$)-axis of the CST crystal generates nonlinear current ($J_{NL}$), which is proportional to the effective nonlinear conductivity tensor, $\sigma_{eff}$, in frequency domain is given by

$$\vec{J}_{NL} \sim \vec{\sigma}_{eff} I_{pu}(\omega) \tag{2}$$

Here, $I_{pu}(\omega)$ is the Fourier transformation of the optical pulse in time domain, $I_{pu}(t)$. The effective nonlinear conductivity coefficient, $\vec{\sigma}_{eff}$, which in sample coordinate frame can be expressed as (refer to the note 5.1 of SI for details):

$$\vec{\sigma}_{eff} = (\sigma_{11}\cos 2\varphi + \sigma_{16}\sin 2\varphi)\vec{x} + (-\sigma_{16}\cos 2\varphi + \sigma_{11}\sin 2\varphi)\vec{y} + \sigma_{31}\vec{z} \tag{3}$$

where $\varphi$ denotes the azimuthal angle ($\varphi$) or polarization angle ($\alpha$) as illustrated in Fig. 2(c).

When the pump polarization is fixed along $x_0$-axis (horizontal) in laboratory coordinate, the generated THz electric field components $E_H^{THz}$ and $E_V^{THz}$ with respect to the azimuthal angle $\varphi$ is (refer to the note 5.2 of SI for details):

$$E_H^{THz} \sim E_x^{THz}\cos\varphi - E_y^{THz}\sin\varphi \propto \sqrt{\sigma_{11}^2 + \sigma_{16}^2}\sin(3\varphi + \Phi) \tag{4}$$

and $\quad E_V^{THz} \sim E_x^{THz}\sin\varphi + E_y^{THz}\cos\varphi \propto \sqrt{\sigma_{11}^2 + \sigma_{16}^2}\sin\left(3(\varphi + \frac{\pi}{6}) + \Phi\right) \tag{5}$

where $\Phi = \tan^{-1}(\sigma_{11}/\sigma_{16})$. The generated THz $E$-field, $E_{THz}$, as a function of azimuthal angle $\varphi$ is proportional to $\sin(3\varphi+\Phi)$. By combining Eqs. (4) and (5), the phase difference between $E_H$ and $E_V$ in each period is easily found to be 30°. The solid lines in Fig. 4(a) are the fitting curves with the sine function, which produce the phase difference of 33.6° between the two curves, close to the theoretical value predicted by Eqs. (4) and (5).

When sample position is fixed, i.e. $\theta$ remains unchanged, the incident polarization $E$ of optical pulse is varied by rotating a half-wave plate. The horizontal and vertical components of the generated THz electric field are (refer to the note 5.3 in SI for details):

$$E_H^{THz} \sim E_x^{THz}\cos\theta - E_y^{THz}\sin\theta \propto \sqrt{\sigma_{11}^2 + \sigma_{16}^2}\sin(2\alpha + \theta + \Phi) \tag{6}$$

and

$$E_V^{THz} \sim E_x^{THz}\sin\theta + E_y^{THz}\cos\theta \propto \sqrt{\sigma_{11}^2 + \sigma_{16}^2}\sin\left(2(\alpha + \frac{\pi}{4}) + \theta + \Phi\right) \tag{7}$$

with $\Phi = \tan^{-1}(\sigma_{11}/\sigma_{16})$. It is clearly seen from Eqs. (6) and (7) that the magnitude of THz field changes periodically with $\alpha$ in the relation of $\sin(2\alpha+\theta+\Phi)$. From Eqs. (6) and (7), it is also clear that the phase difference between horizontal and vertical directions is $\pi/4$ in each period. Solid lines in Figure 4(d) are the fitting curves with the sine function, revealing a good agreement between experimental data and simulation. The deviation from the theoretical prediction may come from the roughness of the sample surface due to the larger laser beam used during the measurement.

As discussed above, ultrafast resonant excitation in CST surface can generate transient photocurrent, which can lead to ultrafast THz radiation. Considering our experimental configuration as well as the azimuthal and polarization angle dependence of THz amplitude, we can conclude that only nonlinear photocurrent can contribute to the observed THz radiation. Here, the nonlinear photocurrent has two sources, i.e. the injection current and shift current, both of which are related to the second order nonlinear conductivity.

**Injection current:** The injection current has been referred as the CPGE, which arises when initial and final states of the perturbed electrons have different band velocities with optical excitation. [35-37] The contribution of injection current results from an asymmetry distribution of carriers from momentum space due to the interference between absorption processes induced by orthogonally polarized light, and the maxima current occurs for circularly polarized light. [36] Here we employ the simplified model proposed in refs. [37-39], in which the injection current, $J_{inj}$, induced by ultrafast optical pulse is proportional to the average velocity change ($\Delta v$), and the excited electron density ($n$), and the injection current could be expressed as

$$J_{inj} = c_{inj}\Delta v \int S(t) e^{-t/\tau_{inj}} I_p(t-\tau) d\tau \tag{8}$$

where the $c_{inj}=en\Delta z$ is the sheet charge density with $\Delta z$ being the thickness of the emitting sheet. For simplicity, as done in ref. [37], an exponential term with relaxation time of $\tau_{inj}$ is introduced phenomenological for describing the response of injection current, the $\tau_{inj}$ denotes the electron momentum relaxation time, such as phonon-, and carrier-induced scattering processes. $S(t)$ and $I_p(t)$ are the unit step function and pump pulse, respectively. It is seen that the injection current temporally follows the envelope of pump pulse, $I_p$, the current consequently is expected to show a unipolar following excitation of pump pulse with Gaussian distribution. The THz radiation with circularly polarized pump pulse shown in Figure 3(d) indicates injection currents play a dominated role for the THz emission, in which the opposite helicity of optical pulse gives rise to THz signal with π phase shift. The photocurrent can be obtained by the integration of THz signal of Fig. 3(d) with time. As the THz signal with circularly polarized pump has poor signal-to-noise ratio, we cannot obtain clearly transient photocurrent. Figure 5(b) shows the pump helicity-dependent THz amplitude, in which the pump polarization undergoes linear, elliptical, right circular polarization, and back to linear when the QWP angle is rotated from 0° to 90°, and more data are given in Figure S5 of SI. It is more clearly that the THz emission shows much larger amplitude for linearly pump pulse, in which case shift current plays a dominated role for THz generation.

**Shift current**: After above bandgap excitation, the LPGE in non-inversion symmetry semiconductor can generate transient photocurrent, [37-40] which involves the shift in center of electron charge induced by absorption of ultrafast optical pulse. The shift is on the order of bondlength, and occurs on fs time scales. [38] For short pulse excitation, this process leads to a step-like charge displacement $\Delta x_{sh} S(t)$ whose temporal derivative is proportional to the shift current $J_{sh}$. By introducing an exponential term representing the relaxation time of shift current, $\tau_{sh}$, we have [37]

$$J_{sh} = c_{sh} \frac{\partial}{\partial t} \int \Delta x S(t) e^{-t/\tau_{sh}} I_p(t-\tau) d\tau \qquad (9)$$

Where $c_{sh}=en\Delta z_{sh}$, this model implies $J_{sh}$ initially follows the profile of $I_p(t)$ and becomes bipolar if the relaxation time $\tau_{sh}$ is comparable to or longer than the pump duration. [37-39] Figure 5(a) is obtained by integrating the THz electric field of black curve in Figure 3(d) under linearly polarized excitation, see the details in note 7 of SI. We can see that the relaxation time of shift current in our sample is comparable or longer than 120 fs, the duration of the optical pulse we used, according to our experimental results.

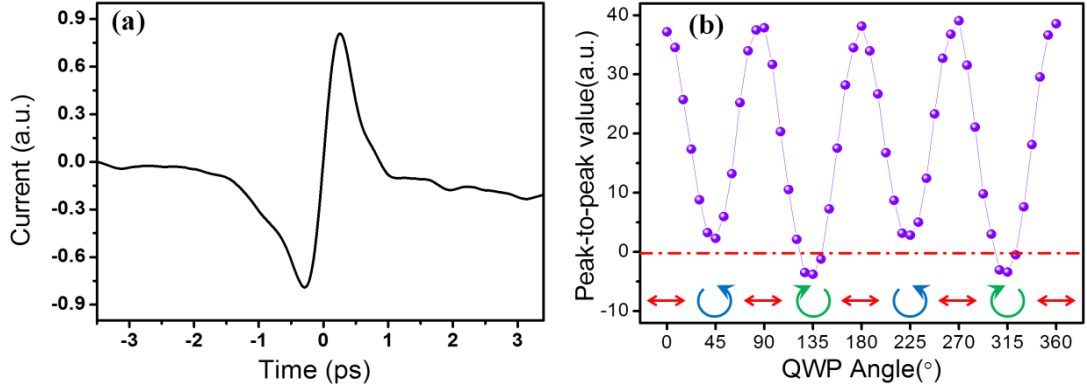

Figure 5 (a) The photocurrent obtained by integrating the black curve in Fig. 3(d), i.e. the integration on the THz electric field under linearly polarized pump. (b) Pump polarization dependent peaking amplitude of the emitted THz radiation. The dot-dash line marks the zero amplitude of THz field, guidance for eyes.

Considering that the used photon energy of 1.55 eV in our experiment is even larger than the size of direct band gap, 1.2 eV, not to mention the size of indirect band gap, 0.4 eV, the THz emission due to surface current can arise from both the optical excitation and the subsequent relaxation processes. In order to make a distinction between these two processes, we performed optical pump and THz probe experiments.[41] If the relaxation processes contribute to the THz emission, it is expected that transient dynamics of photocarrier should display ultrafast response with subpicosecond lifetime. Figure 6(a) presents the transient THz transmission under various pump fluences, which shows that the dynamics of relaxation could be well reproduced with bi-exponential function. The inset in Figure 6(a) plots the peak transmission change at zero delay time, i.e. $(\Delta T/T_0)_{t=0}$, with respect to the pump fluence, which displays a good linear relationship, that indicates no saturable effect until the highest pump fluence of 482 $\mu J/cm^2$. With bi-exponential fitting, Figure 6(b) displays the pump fluence dependent fitting lifetimes. It is clear that the fast component increases slightly with pump fluence, in which the magnitude of $\tau_1$ increases from 3.0 ps at 121 μJ/cm$^2$ to 4.0 ps at 482 μJ/cm$^2$. The slow component has a constant lifetime of $\tau_2=23\pm3$ ps, nearly independent of the pump fluence. We assigned tentatively the fast relaxation as the electron-phonon coupling and the slow one as the surface phonon/defect-mediated electron-hole recombination process. The absence of the subpicosecond relaxation process thus could exclude the subsequent photocarrier relaxation as the origin of the observed THz emission and confirm the dominated role of the shift photocurrent generated during optical excitation.

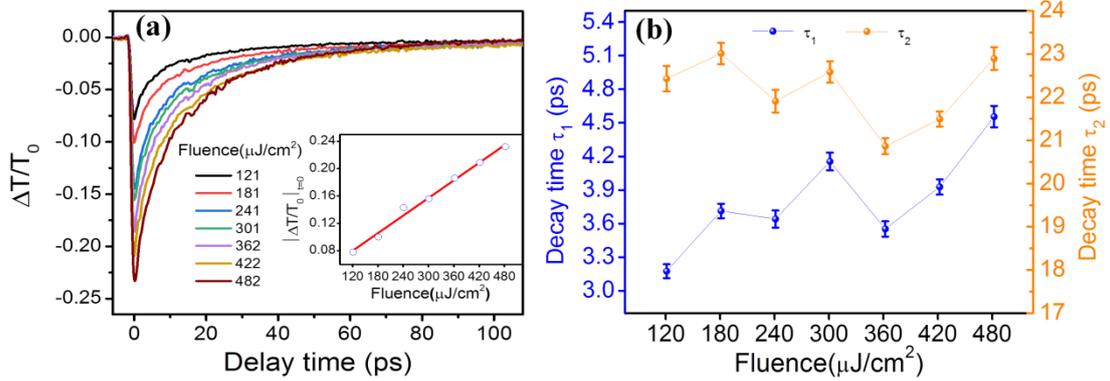

**Figure 6. Transient THz transmission in CrSiTe$_3$**. (a) Transient THz transmission under various pump fluences at a wavelength of 800 nm. The inset shows magnitude of $|\Delta T/T_0|$ at zero delay time with respect to pump fluence. (b) The magnitudes of fitting time constant ($\tau_1$ and $\tau_2$) under various pump fluences.

To summarize, by performing fs laser pulse illumination on the high-quality CrSiTe$_3$ single crystal, a broad band THz emission is observed, which increases linearly in strength with increasing the pump pulse intensity. Considering the inversion symmetry of the bulk crystal, the strong THz emission can only come from the surface with thickness of a few QLs of the crystal, in which the inversion symmetry is broken. Our analysis on the experimental data suggests that the THz emission is dominated by the shift photocurrent occurring on the crystal surface, and the injection current contribution to THz emission is one order of magnitude weaker than that of shift current. Furthermore, optical pump and THz probe experiments reveal that the photocarrier relaxation follows two processes. The fast process with lifetime of a few ps increases slightly with the pump fluence, which is assigned as the electron-phonon coupling process. The slow process with lifetime of 23±3 ps comes from the phonon/defect-mediated electron-hole recombination. The ps relaxation process reveals that the ultrafast THz emission takes place during the optical excitation rather than the subsequent relaxation process. Our experimental results not only have demonstrated for the first time the strong THz emission on the surface of the $CrSiTe_3$ crystal with thickness of a few QLs, but also have showed ultrafast nonlinear optical response in this type of bulk layered 2D magnet. Additionally, the present experimental findings indicate that this kind of 2D magnet provides a good candidate for THz emitter with high conversion efficiency, which makes the THz spectroscopy a sensitive tool for probing the surface/interface of the few-layer and/or monolayer 2D magnets.

**Experimental Section**

*$CrSiTe_3$ single crystal growth*. The $CrSiTe_3$ single crystals were grown using a self-flux method. Starting materials of Cr powder (99.999% Alfa Aesar), Si (99.999%, Alfa Aesar) and Te (99.9999%, Alfa Aesar) blocks thoroughly mixed in a molar ratio of 1:2:6 were placed into an alumina crucible. The crucible was then sealed in a quartz tube in vacuum and was heated up to 1150 °C within 15 hrs in a furnace, held at this temperature for 20 hrs, and subsequently cooled down to 800 °C at a rate of 1.5 °C/h. After staying at 800 °C for more than 5 hours, the excess Si and Te was quickly removed at this temperature in a centrifuge and black crystals with shining surface were finally left. The as-grown crystals are plate-like with the in-plane dimension of ~ 8 mm in diameter.

*Optical pump and THz probe spectroscopy.* Figure 2(a) schematically shows the experimental arrangement of optical pump THz probe spectroscopy. The laser beam delivered from Ti: sapphire laser, with a center wavelength of 800 nm, repetition rate of 1 kHz and pulse duration of 120 fs, is split into three beams. The first one is used for ultrafast optical pump. The THz pulses, co-propagating with the pump pulse, is generated by optical rectification and detected by electro-optic sampling in a pair of (110)-oriented ZnTe crystals with the thickness of 1.0 mm, by the other two beams. The spot size of the THz beam on the sample position is 2.0 mm, whereas the spot size of the pump beam on the sample is 6.5 mm, and the large pump spot size ensures a relatively uniform photoexcited region for THz probe.

*THz emission spectroscopy.* THz emission spectroscopy is similar as that of Figure 2(a) with the pump beam was blocked and the emitted ZnTe was replaced by CrSiTe$_3$ crystal. As shown schematically in Figure 2(b), the *c*-cut CrSiTe$_3$ crystal is irradiated at normal incidence with weakly focused beam with spot size of 2.5 mm. The 120-fs-pulse-duration is much shorter than the period of the emitted THz bursts, enabling the coherent emission. The THz detection system is based on electro-optic sampling with a 1 mm thick (110)-oriented ZnTe crystal. As illustrated in Fig. 2(b), the CrSiTe$_3$ crystal was mounted onto a stage that can rotate in vertical plane for adjusting azimuthal angle ($\varphi$). A half-wave plate was placed before the CrSiTe$_3$ crystal, which is used to change the polarization angle ($\alpha$) of incident laser pulse (see Figure 2(c)). A THz polarizer P was placed after the sample and before the parabolic mirrors PM3, as illustrated in Fig. 2(a), to measure the horizontal and vertical polarization of the emitted THz radiation, respectively.

**Supporting Information**

Supporting Information is available from the Wiley Online Library or from the author.

**Data availability.** The data that support the plots within this paper and other findings of this study are available from the corresponding authors upon reasonable request.

**Acknowledgements:** This work is supported by the National Natural Science Foundation of China (NSFC, Nos. 11674213, 11604202, 61735010, 11874264), the Natural Science Foundation of Shanghai (Grant No. 17ZR1443300). Z.J. thanks Shanghai Municipal Education Commission (Young Eastern Scholar QD2015020), Science and Technology Commission of Shanghai Municipality (Shanghai Rising-Star Program 18QA1401700). Y.F.G. acknowledges the support by the ShanghaiTech University startup fund.

**Conflict of Interest**

The authors declare no conflict of interest.

## 1. The calculated band structure and density of states for CST single crystal

All first-principle electronic structure calculations were performed using the Vienna *ab initio* simulation package (VASP), with the projector augmented wave (PAW) method of Perdew–Burke–Ernzerhof (PBE) functional to treat the interactions between the valence-electrons and their inner core. A plane-wave basis set along with energy cutoff of 450 eV is used to describe electron wavefunctions. A $4\times4\times1$ Monkhorst-Pack grid has been employed to sample the Brillouin zone. During the simulations, both the lattice constants and positions of all atoms are relaxed until the force is less than 0.01 eV $\text{Å}^{-1}$. The criterion for the total energy is set as $1\times10^{-6}$ eV. To confirm the validity of our method, the lattice constants of bulk CST are optimized to be $a = 6.837$ Å, $c = 20.599$ Å. Calculated crystallographic properties obtained by relaxing the structures are in good agreement with previous reports. [1]

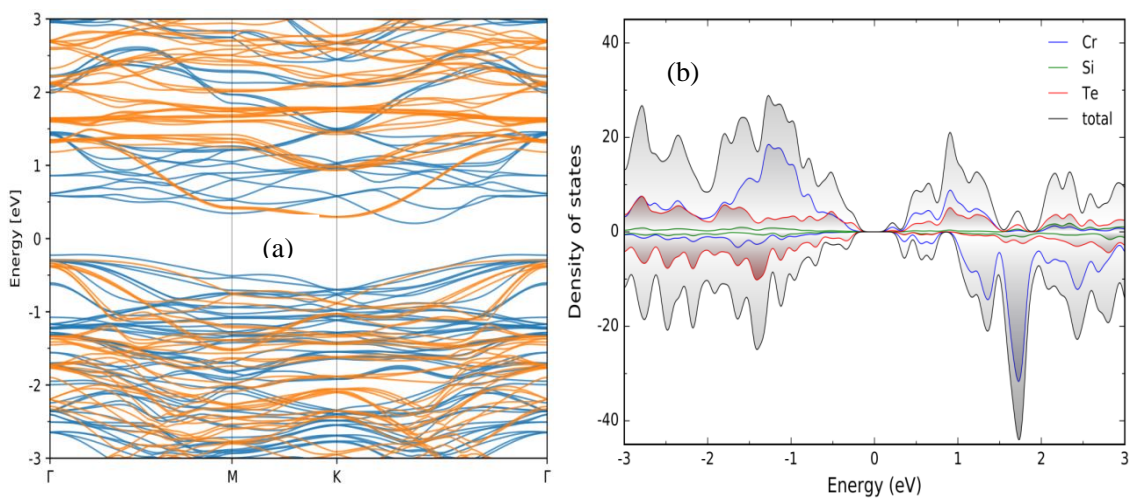

Figure S1. The calculated band structure (a) and density of states (b) of CST single crystal with VASP.

## 2. Refractive index of CST in THz frequency and the temperature dependent THz transmission

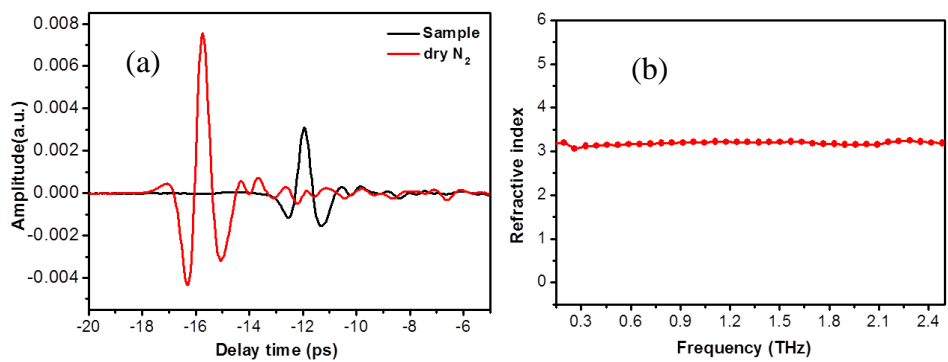

Figure S2.1. (a) THz time domain transmission spectra of dry nitrogen and CrSiTe$_3$ single crystal with thickness of 0.43 mm. (b) the refractive index dispersion of CrSiTe$_3$ crystal in frequency range from 0.2 to 2.5 THz, and the average index of fraction is about 3.17.

In addition, we also measured the transmitted THz polarization change with respect to the azimuthal angle, which is presented in Figure S.2.2. It is clearly seen that the THz polarization does not show any observable change with rotating the sample's azimuthal angle from 0 to 180°, indicating that the property of the *ab*-plane for the *c*-cut CST crystal is isotropic.

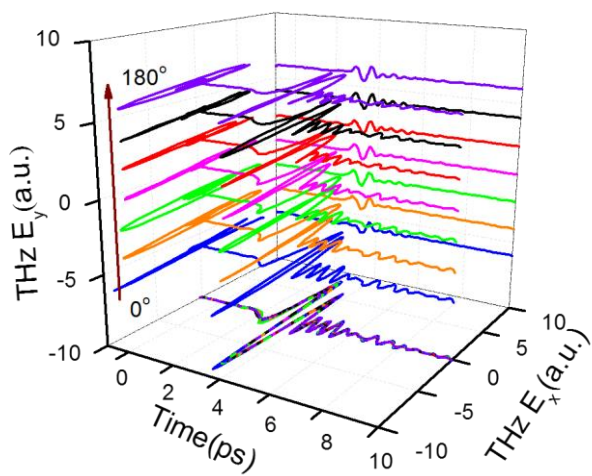

Figure S2.2. A 3D plot of THz transmission of CST crystal with respect to the azimuthal angle at room temperature.

**Temperature dependent THz transmission in CST crystal.**

It is noted that the CST crystal undergoes a ferromagnetic ordering around $T_C = 33$ K [2]. In order to exclude the influence of phase transition on the THz transmission spectra, Figure S2.3 shows the temperature dependent THz transmission spectra, it is clear that the THz transmission is nearly temperature independence in temperature range of 5 - 300 K.

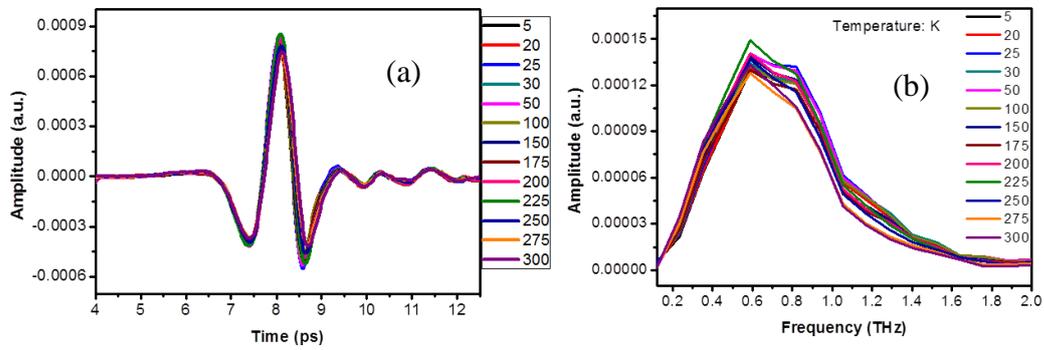

Figure S2.3 THz transmission in time domain (a) and in frequency domain (b) with Fourier transformation in temperature range from 5 K to 300 K.

## 3. The 3D plots of emitted THz polarization with respect to the azimuthal angle

Both horizontal ($E_H$) and vertical ($E_V$) polarization changes of the THz emission with respect to the azimuthal angle can be evaluated by using a THz polarizer. With rotating the c-cut CST crystal around z-axis with angle $\varphi$, Figure S3 presents the 3D plots of the THz radiation.

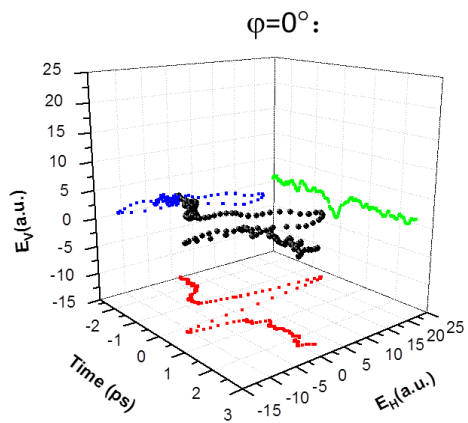
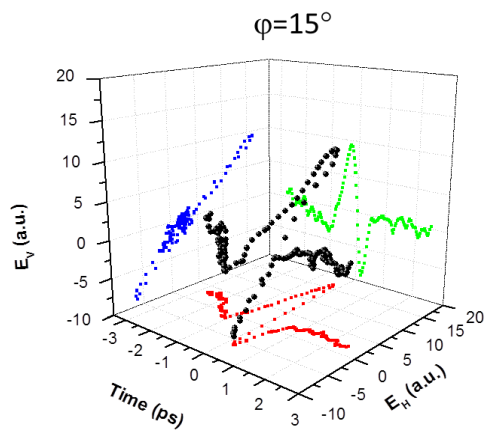
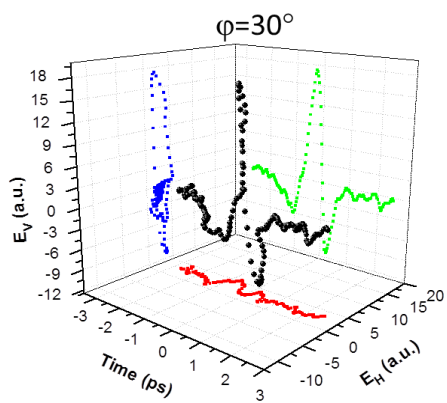
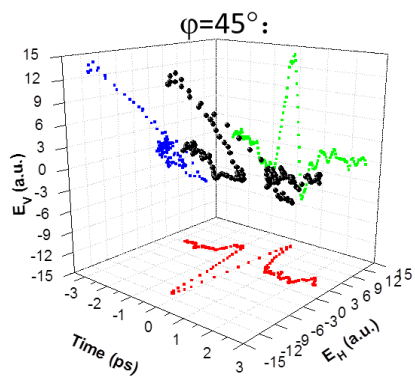
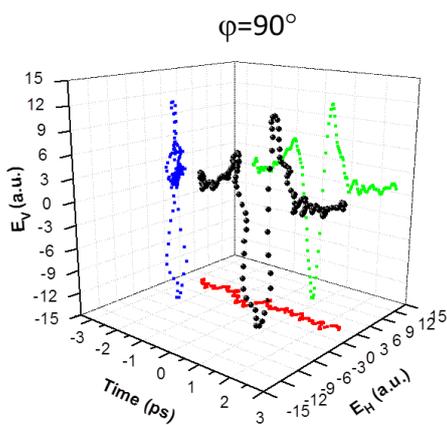
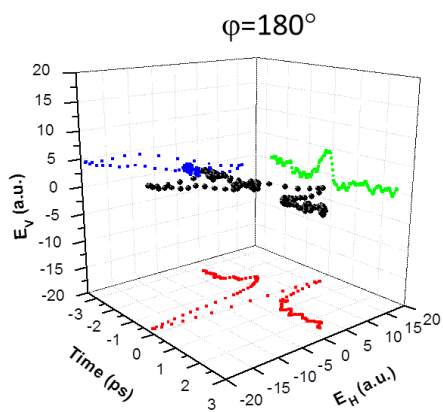
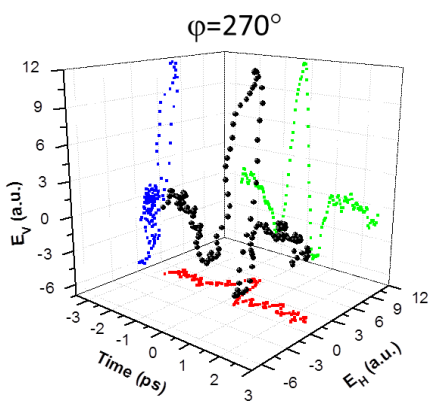
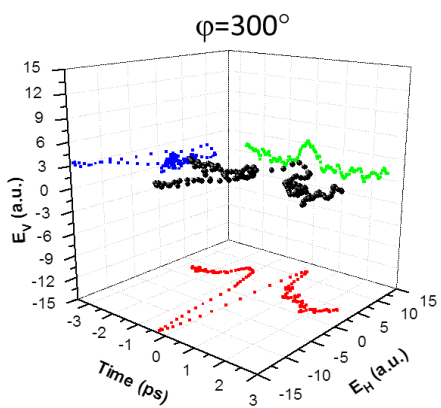

Figure S3. The 3D plots of the polarization of the THz emission with respect to the azimuthal angle $\varphi$.

## 4. The 3D plot of emitted THz polarization with respect to the pump polarization angle

When the sample position is fixed, we changed the pump polarization by rotating a half wave plate in optical path of pump beam, so that both $E_H$ and $E_V$ polarization changes of the THz emission with respect to the polarization angle is obtained by using a THz polarizer. With changing the pump polarization angle $\alpha$, Figure S4 presents the 3D plots of the THz radiation.

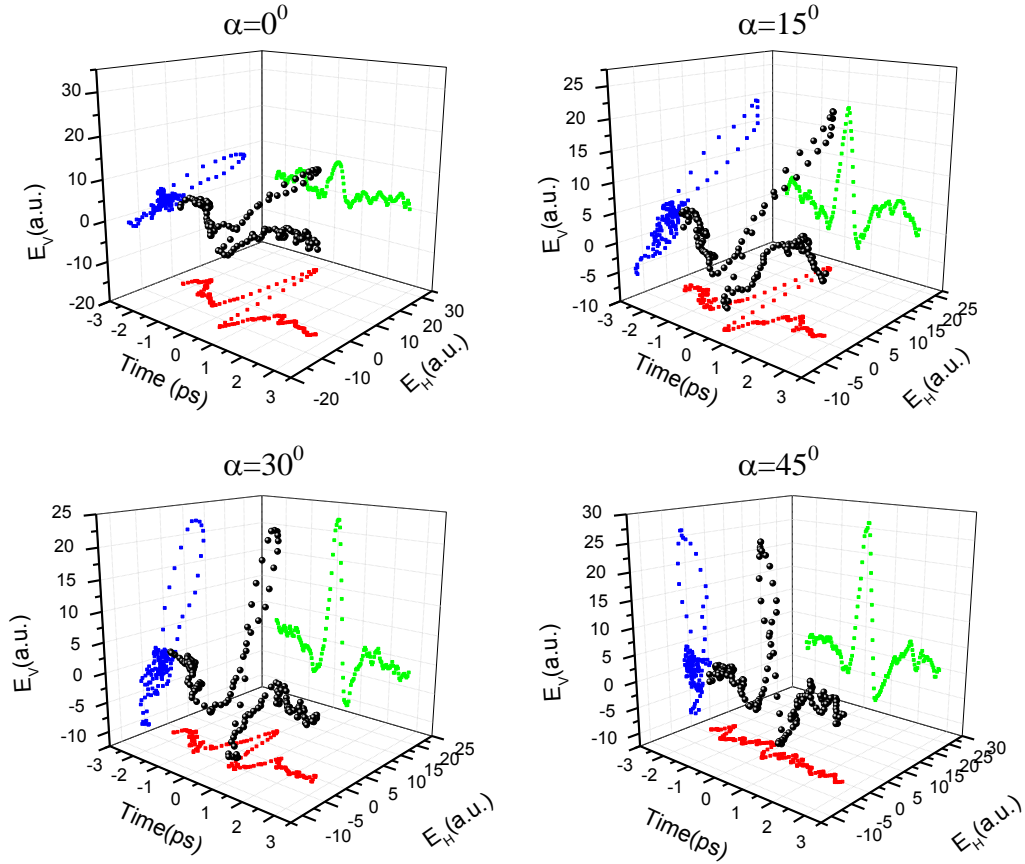

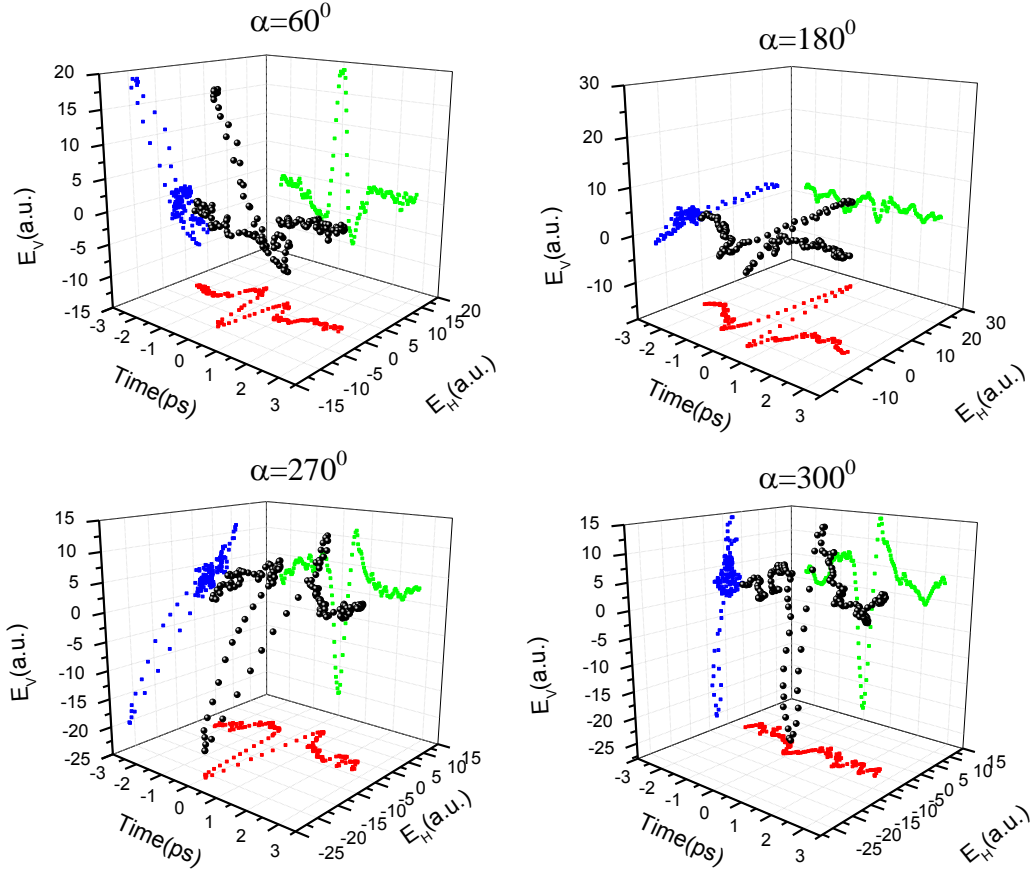

Figure S4. The 3D plots of the polarization of the THz emission with respect to the polarization angle $\alpha$.

## 5. Details for obtaining the effective nonlinear conductivity.

As shown in the inset of Fig. 3(c), the magnitude of THz signal is proportional to the square of the pump field, and the photoinduced current density, J, can be phenomenologically be described by

$$J_{i,NL}(r,t) \sim \sum_{j,k} \iiint \int dt_1 dt_2 d^3r_1 d^3r_2 \sigma_{ijk}(t-t_1, t-t_2, r-r_1, r-r_2) E_j(r_1,t_1) E_k(r_2,r_2) \quad (S1)$$

Where $E_j$ is the $j^{th}$ Cartesian component of the pump field, $\sigma_{ijk}$ is the third-rank tensor describing the nonlinear conductivity of the material. Applying Fourier transformation, the current density can be transformed into frequency domain by

$$\mathbf{J}(\mathbf{k},\Omega) \sim \sigma_{ijk} \mathbf{E}_j(\mathbf{k}_1,\omega_1) \mathbf{E}_k(\mathbf{k}_2,\omega_2) \quad (S2)$$

For dipole approximation, the both **J** and **E** are independent of wavevector **k**, and the Eq. (3) can be reduced to

$$J(\Omega) \sim \sigma_{ijk} E_j(\omega_1) E_k(\omega_2) \quad (S3)$$

**5.1 Derivation of the effective nonlinear conductivity $\sigma_{eff}$,**

The single crystal CST can be indexed as $R\bar{3}$ space group with a $C_{3i}$ point group, the inversion symmetry is broken at the surface/interface of the sample, the surface structure is then reduced to $C_3$ point group, in which the second order nonlinear conductivity, $\sigma_{ijk}$-matrix, in analogue to the second order susceptibility d-matrix, can be written as:

$$\begin{pmatrix} \sigma_{11} & -\sigma_{11} & 0 & \sigma_{14} & \sigma_{15} & -\sigma_{16} \\ -\sigma_{16} & \sigma_{16} & 0 & \sigma_{15} & -\sigma_{14} & -\sigma_{11} \\ \sigma_{31} & \sigma_{31} & \sigma_{33} & 0 & 0 & 0 \end{pmatrix} \quad (S4)$$

The femtosecond laser pulse is incident along *c*-axis of the crystal, $\varphi$ is the azimuthal angle, defined as the angle between the optical electric field and the x-axis of the crystal, and the optical field $\vec{E}$ can be written as a 1×3 matrix,

$$\vec{E} = E_0 \begin{pmatrix} \cos\varphi \\ -\sin\varphi \\ 0 \end{pmatrix} \quad (S5)$$

The pump laser pulse is written as $I_{pu}(\omega) = E_0^2 \exp(-(\omega-\omega_0)^2/\Gamma^2)$, with $E_0$ and $\omega_0$ representing the amplitude as well as the central frequency $\omega_0 = 2.36 \times 10^{15}$ rad/s ($\lambda_0 = 800$ nm) of incident laser pulse. Then the second order nonlinear photocurrent $J$ can be expresses as:

$$\begin{pmatrix} J_x \\ J_y \\ J_z \end{pmatrix} \sim \begin{pmatrix} \sigma_{11} & -\sigma_{11} & 0 & \sigma_{14} & \sigma_{15} & -\sigma_{16} \\ -\sigma_{16} & \sigma_{16} & 0 & \sigma_{15} & -\sigma_{14} & -\sigma_{11} \\ \sigma_{31} & \sigma_{31} & \sigma_{33} & 0 & 0 & 0 \end{pmatrix} \begin{pmatrix} \sin^2\varphi \\ \cos^2\varphi \\ 0 \\ 0 \\ 0 \\ -2\sin\varphi\cos\varphi \end{pmatrix}$$

$$= \begin{pmatrix} -\sigma_{11}\cos 2\varphi - \sigma_{16}\sin 2\varphi \\ \sigma_{16}\cos 2\varphi - \sigma_{11}\sin 2\varphi \\ \sigma_{31} \end{pmatrix} \qquad (S6)$$

Then, the effective nonlinear coefficient in sample coordinate system, $\sigma_{\text{eff}}$ is then obtained as

$$\vec{\sigma}_{\text{eff}} = (\sigma_{11}\cos 2\varphi + \sigma_{16}\sin 2\varphi)\vec{x} + (-\sigma_{16}\cos 2\varphi + \sigma_{11}\sin 2\varphi)\vec{y} + \sigma_{31}\vec{z} \qquad (S7)$$

Nonlinear current J is linear proportional to $\sigma$, and the emitted THz electric field $E_{\text{THz}}$ is proportional to $\partial J/\partial t$, it is obvious that the magnitude of $E_{\text{THz}}$ is therefore proportional to the magnitude of $\sigma$.

### 5.2 Derivation of azimuthal angle dependent $E_H$ and $E_V$ component of THz radiation

Here, the polarization of optical pulse is set horizontally, therefore we have $\varphi = \theta$. The $E_H$ and $E_V$ component of THz radiation then is

$$\begin{aligned} E_H &= E_x\cos\theta - E_y\sin\theta = E_x\cos\varphi - E_y\sin\varphi \\ &\sim \left[(\sigma_{11}\cos 2\varphi + \sigma_{16}\sin 2\varphi)\cos\varphi - (-\sigma_{16}\cos 2\varphi + \sigma_{11}\sin 2\varphi)\sin\varphi\right] \\ &= \left[\sigma_{11}\cos 2\varphi\cos\varphi + \sigma_{16}\sin 2\varphi\cos\varphi + \sigma_{16}\cos 2\varphi\sin\varphi - \sigma_{11}\sin 2\varphi\sin\varphi\right] \\ &= \sqrt{\sigma_{11}^2 + \sigma_{16}^2}\sin(3\varphi + \Phi) \end{aligned} \qquad (S8)$$

$$\begin{aligned} E_V &= E_x\sin\theta + E_y\cos\theta = E_x\sin\varphi + E_y\cos\varphi \\ &\sim \left[(\sigma_{11}\cos 2\varphi + \sigma_{16}\sin 2\varphi)\sin\varphi + (-\sigma_{16}\cos 2\varphi + \sigma_{11}\sin 2\varphi)\cos\varphi\right] \\ &= \left[\sigma_{11}\cos 2\varphi\sin\varphi + \sigma_{16}\sin 2\varphi\sin\varphi - \sigma_{16}\cos 2\varphi\cos\varphi + \sigma_{11}\sin 2\varphi\cos\varphi\right] \\ &= \sqrt{\sigma_{11}^2 + \sigma_{16}^2}\sin\left(3(\varphi + \frac{\pi}{6}) + \Phi\right) \end{aligned} \qquad (S9)$$

where $\Phi = \tan^{-1}(\sigma_{11}/\sigma_{16})$.

Obviously, the emitted THz E-field shows azimuthal angle dependence with periodicity of 120° shown in Eq. (S8), where the phase retardation between $E_V$ and $E_H$ is 30° as given in Eq.(S9)

### 5.3 Derivation of pump polarization dependent $E_H$ and $E_V$ components of THz emission

The included angle $\alpha$ denotes the angle between the the polarization of pump pulse and the x-axis of the crystal, which is changed by a half-wave plate. As illustrated in Figure 2(c), the angle $\theta$ is fixed at this situation. The $E_H$ and $E_V$ component of THz radiation is then,

$$\begin{aligned}
E_H &= E_x \cos\theta - E_y \sin\theta \\
&\propto (\sigma_{11}\cos 2\alpha + \sigma_{16}\sin 2\alpha)\cos\theta - (-\sigma_{16}\cos 2\alpha + \sigma_{11}\sin 2\alpha)\sin\theta \\
&\propto \sigma_{11}(\cos 2\alpha \cos\theta - \sin 2\alpha \sin\theta) + \sigma_{16}(\sin 2\alpha \cos\theta + \cos 2\alpha \sin\theta) \\
&\propto \sqrt{\sigma_{11}^2 + \sigma_{16}^2}\sin(2\alpha + \theta + \Phi)
\end{aligned} \quad (S10)$$

$$\begin{aligned}
E_V &= E_x \sin\theta + E_y \cos\theta \\
&\propto (\sigma_{11}\cos 2\alpha + \sigma_{16}\sin 2\alpha)\sin\theta + (-\sigma_{16}\cos 2\alpha + \sigma_{11}\sin 2\alpha)\cos\theta \\
&= \sigma_{11}\sin(2\alpha + \theta) - \sigma_{16}\cos(2\alpha + \theta) \\
&= \sqrt{\sigma_{11}^2 + \sigma_{16}^2}\sin(2(\alpha + \frac{\pi}{4}) + \theta + \Phi)
\end{aligned} \quad (S11)$$

where $\Phi = \tan^{-1}(\sigma_{11}/\sigma_{16})$.

Eqs.(S10) and (S11) demonstrate that polarization dependence of THz E-field shows a periodicity of 180° shown and the phase retardation between $E_V$ and $E_H$ is 45°, respectively.

### 6  Pump helicity-dependent THz emission

Figure 5(b) in the main text is extracted from the Figure S5 below, in which the QWP is the abbreviation of quarter-wave plate

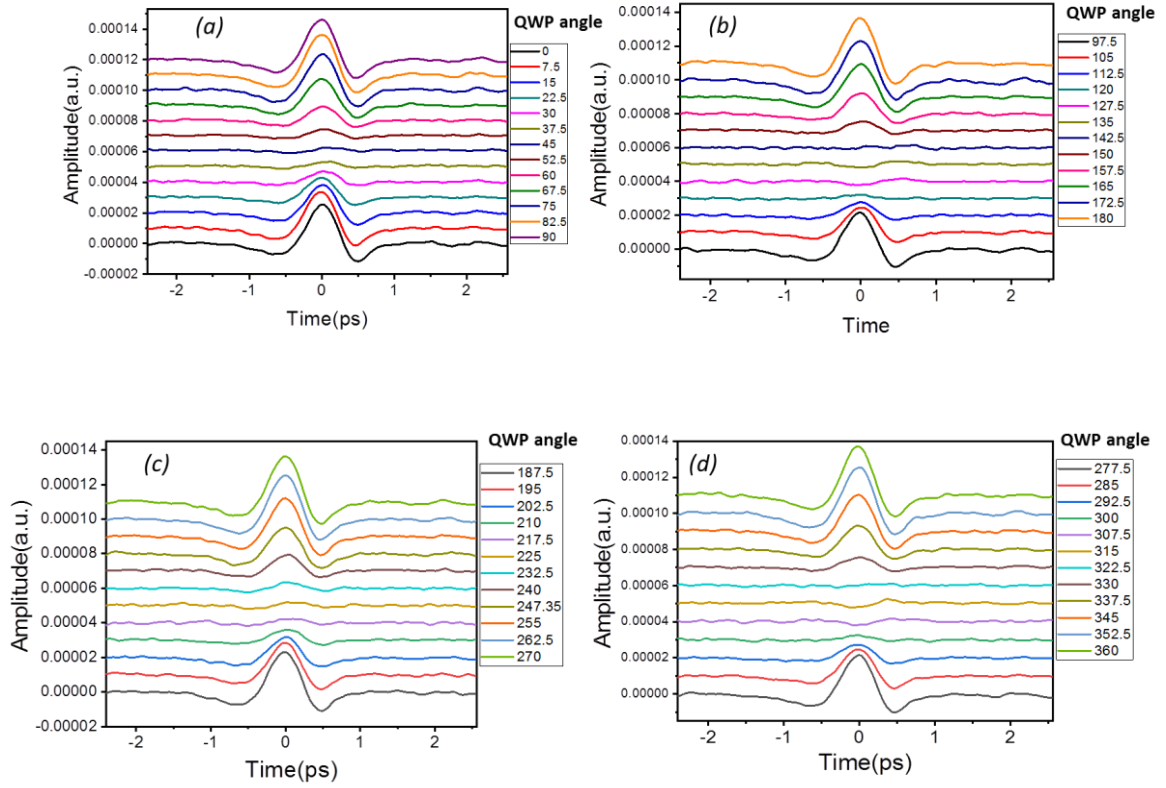

Figure S5 Pump helicity dependence of THz emission. The quarter wave plate angle is rotated from 0 ° to 360 ° with step of 7.5 °.

## 7  Evaluation of THz current from THz fields

In order to obtain the THz currents $J(t)$ from the measured far field THz electric field $E(t)$, we applied Fraunhofer approximation, which yields that the peak electric field $E(t)$ is proportional to the derivative of source current, $J(t)$. For the case of normal incidence with transmission configuration, the relationship between the $E(t)$ and $J(t)$ can be expressed as [3]:

$$E(t) \sim \frac{1}{R}\frac{Z_0}{1+n}\frac{\partial J(t)}{\partial t} \qquad (S12)$$

$Z_0 = 377$ is the vacuum impendence, $n = 3.17$ is the refractive index of the CST crystal. $R$ denotes the distance between the sample and EO sampling crystal. The $J(t)$ denotes photocurrent density, the quantiy $\partial J(t)/\partial t$ is evaluated at the retarded time $t` = t - R/c$. We would like to mention that the THz electric field $E(\omega)$ is related to the measured signal $S(\omega)$ which is obtained from the fast Fourier transformation of the measured signal $S(t)$) on the EO sampling crystal via $S(\omega)=h(\omega)E(\omega)$, where the transfer function $h(\omega)$ is supposed to be a constant, $b$, in our case. We therefore have $S(\omega) = bE(\omega)$. By applying inverse fast Fourier transformation, we can obtain the far field THz electric field, $E(t)$. By integrating the $E(t)$ with time, the source current $J(t)$ can be evaluated, which is displayed in Figure 5(a).